\documentclass[12pt]{scrartcl}
\usepackage[english,russian]{babel}
\usepackage{graphicx}
\usepackage{mathtext}
\usepackage{indentfirst}
\usepackage{epsfig,amsmath,amsfonts}

\usepackage{hyperref}
\def\n{\nu}    
    
\def\ff{\phi}   

  \def\G{\Gamma}

\def\fr{\frac} \def\dfr{\dfrac} \def\dt{\partial}

\def\beq{\begin{equation}}
\def\eeq{\end{equation}}

\def\bear{\begin{eqnarray}}
\def\eear{\end{eqnarray}}

\def\bea*{\begin{eqnarray*}}
\def\eea*{\end{eqnarray*}}

\def\ph{\phantom}
\def\mc{\mathcal}

\begin{document}

\renewcommand{\contentsname}{}
\renewcommand{\refname}{\begin{center}References\end{center}}
\renewcommand{\abstractname}{\begin{center}\footnotesize{\bf Abstract}\end{center}}

\hfill\textsc{QMUL-PH-11-11}

\vspace{3cm}

\begin{center}
\textsc{\Large Boundary Terms in Generalized Geometry and doubled field theory}

\medskip
\large{David S. Berman${}^1$, Edvard T. Musaev${}^1$, Malcolm J. Perry${}^2$}
\medskip

{\small ${}^1$  \it  Department of Physics, Queen Mary University of London, Mile End Road, London}

{\small ${}^2$  \it Centre for Mathemetical Science, DAMTP, Wilberforce Road, Cambridge}

\end{center}

\abstractname{}
{\abstract We propose a boundary action to complement the recently developed duality manifest actions in string and
M--theory using  generalized geometry. This boundary action combines the Gibbons--Hawking term with boundary pieces that
were previously neglected in the construction of these actions. The combination may be written in terms of the metric of
generalized geometry. The result is to produce an action that is duality invariant including boundary terms.  }

\section{Introduction}

Dualities have always played a central role in string theory. T--duality is a hidden symmetry from the spacetime point
of view and a nonperturbative symmetry from the world--sheet point of view \cite{Giveon:1994fu}. Already during the
early days of string theory there were attempts to reformulate the theory to make T--duality a manifest symmetry
\cite{Duff:1989tf,Tseytlin:1990nb,Tseytlin:1990va,Maharana:1992my,Siegel:1993th,Siegel:1993xq}. This was done by doubling the dimension of the space where T--duality acts so that T--duality was
linearly realized on this space. The metric on this doubled space turns out to be the same as that of the generalized geometry
introduced by Hitchin \cite{Hitchin:2010qz}. A useful and natural by product is the incorporation of the NS--NS two
form potential into the generalized metric so that one only needs the generalized metric on the doubled space and the
dilaton. The dilaton is not doubled but is shifted from the usual string frame dilaton; this is related to how the
dilaton is required to shift under T-duality. (In doubled field theory the Ramond Ramond sector must be incorporated seperately see \cite{Hohm:2011dv,Coimbra:2011nw} for recent work in this direction.)

Recently the doubled approach has had something of a rebirth \cite{Hull:2004in,Hull:2009mi}. Its quantum properties have been investigated in \cite{Berman:2007vi,Berman:2007xn,Berman:2007yf,Copland:2011yh}. It has also been extended to M--theory in
\cite{Hull:2007zu, Pacheco:2008ps, Berman:2010is,Berman:2011pe, Hillmann:2009ci} where the  duality group is related to the U--duality groups of string theory. (For a review of U-duality see \cite{Obers:1998fb}.) The relationship between the doubled formalism in string theory and the M-theory extended geometry is explored in detail in \cite{Thompson:2011uw}. The type II string has been developed in \cite{Hohm:2011zr}. All these recent approaches may be viewed as a small part of the larger programme to encode the symmetries of M-theory into $E_{11}$. For early work in this direction see \cite{West:2001as} and more recent work directly related to these ideas see \cite{West:2010rv}.

The use of the generalized metric on the doubled (or, in the case of M--theory, extended) space allows one to construct
a duality manifest action for supergravity in terms of this generalized metric. Importantly, even though the duality was derived using a flat toroidal space, it turns out no such
requirement is necessary for this reformulation of the spacetime action. The duality frame is chosen by simply
specifying how one dimensionally reduces from the whole extended space to physical spacetime. This is discussed in
detail in \cite{Berman:2010is, Berman:2011pe,Hohm:2010pp}.

The duality manifest actions of supergravity will be the topic of this paper. These actions are quadratic in derivatives
whereas the  Einstein-Hilbert action is of course second order in derivatives. Thus, the equivalence of the duality
manifest actions to the usual action was demonstrated only after an integration by parts which turns the Einstein-Hilbert into an action quadratic in first derivatives and then one of course neglects boundary terms. A natural question given generalized geometry defined by a metric is how does one construct a connection and curvature. The attempts to construct these ideas in generalized geometry are described in \cite{Jeon:2010rw,Hohm:2010xe,Jeon:2011cn,Coimbra:2011nw}.

This equivalence (and subsequent neglect of boundary terms) was described for the doubled string in \cite{Hohm:2010pp},
the extended heterotic  string in \cite{Hohm:2011ex} and in M--theory in \cite{Berman:2010is,Berman:2011pe,Hillmann:2009ci}.

There are two possibilities. The boundary action could break the duality manifest form or it can be reformulated in
terms of the  generalized metric and the boundary action itself becomes reformulated in a duality invariant way. Both
possibilities would be allowed logically.

Famously, gravity already contains a boundary action, the York-- Gibbons--Hawking term \cite{York:1972sj,Gibbons:1976ue} (usually just called Gibbons--Hawking). The
necessity for the Gibbons--Hawking term in the action has its origins in deriving the equations of motion from the
Einstein-Hilbert subject to appropriate boundary conditions. This term is also required for completness of transition amplitudes in quantum gravity \cite{Hawking:1980gf}. The definition of the complete gravitational action then
has the rather useful by product of allowing a simple calculation of blackhole thermodynamics. The action when
evaluated on shell in a Euclideanized background becomes equivalent to the free energy of the system. Thermodynamic
quantities such as energy and entropy are then found from the free energy just by taking derivatives. For the
Schwarzchild black hole the bulk action vanishes on shell and all the thermodynamic information is contained in the the
Gibbons-Hawking boundary term.

The point here is to emphasize that actions have applications beyond simply encoding the equations of motion and naive
quantization.  In gravity, they encode the thermodynamics of solutions.

This paper will combine boundary terms arising from two sources the boundary terms, required to equate the reformulated
duality  manifest actions with the Einstein-Hilbert action and the Gibbons-Hawking boundary term. The combination will
then be written in terms of the generalized metric. The equivalence to the usual form will come from restricting the
boundary on the doubled (or extended space). These restrictions on the boundary will be consistent with the dimensional
reduction conditions required to show equivalence of the bulk actions.

Other recent interesting work on doubled field theory is given in \cite{Albertsson:2008gq,Albertsson:2011ux,Kan:2011vg,Aldazabal:2011nj,Jeon:2011vx,Jeon:2011kp}.

\section{Doubled field theory and generalized geometry}

The usual bosonic part of the low energy effective action for NS-NS sector of the conventional
string is as follows:
\begin{equation}
 \label{2.1}
    S=\int_{\mc{M}}  \sqrt{g}e^{-2\ff}\left(R[g]+4(\dt\ff)^2-\fr{1}{12}H^2\right),
\end{equation}
where $H_{ijk}=3\dt_{[i}b_{jk]}$ is a field strength for the Kalb--Ramond field. 

To make the $O(d,d)$ symmetry manifest, the space is doubled by including so called {\it{winding--mode}} coordinates which then allow the action to be written in 
terms of a generalized metric $\mc{H}_{MN}$ \cite{Hull:2009mi} on the doubled space. We denote the doubled space by ${\mc{M}}^*$ with coordinates $X^M=(\tilde{x}_i,x^j)$, where $M=1,..,2d$. The action in terms of the generalized metric on the doubled space is given by:
\begin{equation}
 \label{2.2}	
\begin{split}
    S=\int_{{\mc{M}}^*} e^{-2d}&\left(\fr18\mc{H}^{MN}\dt_M\mc{H}^{KL}\dt_{N}\mc{H}_{KL} -\fr12
\mc{H}^{KL}\dt_L\mc{H}^{MN}\dt_N\mc{H}_{KM}  \right.\\
      & \left.- 2\dt_M d\dt_N\mc{H}^{MN}+4\mc{H}^{MN}\dt_Md\dt_Nd\lefteqn{\ph{\fr12}}\right) \, .
\end{split}
\end{equation}
The doubled dilaton $d$ is written in terms of the usual dilaton $\ff$ as follows
\begin{equation}
e^{-2d}=\sqrt{g}e^{-2\ff}  \, .
\end{equation}
The generalized metric, $\mc{H}_{MN}$ is a quadratic form on this doubled space with coordinates $X^M$ and is constructed from the usual
metric $g_{ij}$ and two form $b_{ij}$ on the nondoubled space as below:
\begin{equation}
\label{2.3}
 \mc{H}_{MN}=
\begin{bmatrix}
   g^{ij}   & b_i^{\ph{i}k} \\
	&\\
   -b^l_{\ph{l}j}		& g_{kl}+b_{k}{}^a b_{la}
\end{bmatrix}, \quad 
X^M= 
\begin{bmatrix}
 \tilde{x}_m \\
  x^m
\end{bmatrix}.
\end{equation}
$\mc{H}^{MN}$ denotes the inverse of $\mc{H}_{MN}$.

The theory is now also equipped with a constraint on the fields that has its origin in the level matching condition. This is given by,
\begin{equation}
\dt_{x^i}\dt_{\tilde{x}_i} {}{} =0 \, .
\end{equation}
This may be written in the doubled $O(d,d)$ invariant form as:
\begin{equation}
\eta^{AB} \dt_A \dt_B  {} {} =0 \, ,
\end{equation}
where $\eta_{AB}$ is the $O(d,d)$ metric:
\begin{equation}
\eta_{AB} = 
\begin{bmatrix}
0 & 1 \\
1 & 0
\end{bmatrix}
\, .
\end{equation}

Different duality frames are chosen by different choices of solution of this condition with the most obvious choices being (though by no means the only ones):
\begin{equation}
\dt_{\tilde{x}} {} {} =0 \, \quad  {\rm{or \, \,  its \, \, dual}} \quad \dt_x {} {}= 0 \, .
\end{equation}
Taking the first choice so that all fields are taken to be independent on the {\it{winding}} coordinates $\tilde{x}_i$, it was shown \cite{Hull:2009mi} that \eqref{2.2} reduces to \eqref{2.1} up to boundary terms. For us it will be important to keep the previously neglected boundary terms. Imposing this constraint but keeping boundary terms we obtain the resulting action:
\begin{equation}
\label{2}
\begin{split}
 S &=\int_{\mc{M}}  \sqrt{g}e^{-2\ff}\left(R[g]+2(\dt\ff)^2-\fr1{12}H^2\right) \\
&-\int_{\mc{M}} \dt_m\left[e^{-2\ff}\sqrt{g}g^{nb}g^{mc}\dt_ng_{bc}-e^{-2\ff
}\sqrt{g} g^ { mc } g^ { nb}\dt_cg_{nb} \right].
\end{split}
\end{equation}
It is natural to then combine the total derivative term in the above with the Gibbons--Hawking term (modified by
dilaton). This boundary term introduced by Gibbons and Hawking in \cite{Gibbons:1976ue} is given by:
\begin{equation}
\label{3}
\begin{split}
 S_{GH}=&2\oint_{\partial{\mc{M}}}  \sqrt{h}e^{-2\ff}K=2\oint_{\partial{\mc{M}}}   \sqrt{h}e^{-2\ff}h^{ab}\left(\dt_an_b-\G_{ab}^mn_m\right)\\
	&=2\oint_{\partial{\mc{M}}}   \sqrt{h}e^{-2\ff}h^{ab}\dt_an_b-\oint_{\partial{M}}  \sqrt{h}e^{-2\ff}h^{ab}h^{mn}(2\dt_a h_{nb} - \dt_nh_{ab})n_m
\end{split}
\end{equation}
where $K=\nabla_i n^i$ is the trace of the second fundamental form for the induced metric on the boundary, $n_a$ is normal on the boundary and $h_{ab}$ is metric on
the boundary.

Comparing \eqref{2} and \eqref{3} (and with the replacement of $g$ by $h$) one obtains:
\begin{equation}
\label{4}
 \int_{\mc{M}}\sqrt{g}e^{-2\ff}\left(R[g]+4(\dt\ff)^2-\fr1{12}H^2\right)+S_{GH}=S+\oint_{\partial{M}}  \sqrt{h}e^{-2\ff}(2h^{ab}n_{a,b}-n^ch^{ab}\dt_b
h_{ac}).
\end{equation}

We now wish to write the boundary term on the right hand side of \eqref{4} in $O(d,d)$--covariant form by recasting it in terms of the generalized metric.
This produces,
\begin{equation}
 \label{2.4}
S_{tot}=S+\oint_{\partial \mc{M}^* }  e^{-2d} \left[2\mc{H}^{AB}\dt_AN_B+N_A\dt_B \mc{H}^{AB}\right]  \, .
\end{equation}
The normal $N_A$ is now the unit normal to the boundary in the doubled space. 
The expression $\eqref{2.4}$ is $O(d,d)$ covariant and should be true in any duality frame. 

In order for this term to match the boundary term in \eqref{4} (after a duality 
frame is chosen to give the usual bulk action) we require that the possible normal vector for the boundary in the doubled space be restricted to the form:
\begin{equation}
 N_A=
\begin{bmatrix}
 0\\ \\n_a
\end{bmatrix},\quad
N^A=
\begin{bmatrix}
 -b^i_{\ph{i}j}n_i \\ \\ n^a
\end{bmatrix}.
\end{equation}
This normal is such that the normalization condition doesn't imply any constraints to the dynamical fields $g_{ij}$ and
$b_{ij}$:
\begin{equation}
 N^{A}N^{B}\mc{H}_{AB}=1 \Longrightarrow n_a n^a=1.
\end{equation}

The fact that the normal is only allowed components along the $x^i$ directions is due to the fact that we chose the particular duality frame where fields are independent of $\tilde{x}_i$. A direct consequence of this is that there
could be no boundary located in $\tilde{x}_i$ as this would break $\tilde{x}_i$  translation invariance. Of course, if we chose the T-dual frame where fields are independent of $x^i$ then we would have to choose the opposite condition on the boundary normal. A natural conjecture is that the general restriction on the boundary normal follows from the constraint which has its origins in the level matching condition so that in general we require that:
\begin{equation}
N^A \eta_{AB} N^B =0  \, .
\end{equation}

There is also a doubled geometry for the heterotic string \cite{Hohm:2011ex}. Now there are also new coordinates denoted by $y_i$ that are dual to the guage fields. A similar story follows with an action written in terms of some metric on the extended space the usual low energy action for the heterotic string following from reducing the theory.

The second order derivatives appear only in the Ricci scalar term term of the effective action. Hence, the additional pieces in the generalized metric for the heterotic string do not affect the form of the boundary term and it remains the same as before ie. equation \eqref{2.4}. The normal $N_A$ will be restricted by the requirement of reproducing the usual action once the conditions, 
\begin{equation}
\partial_{\tilde{x}} = 0 \, , \qquad \partial_y =0
\end{equation}
are chosen so that now similarly to before
\begin{equation}
 N_A=
\begin{bmatrix}
    0 \\ n_a \\ 0
\end{bmatrix}.
\end{equation}

\section{M--theory}

The M--theory version of the doubled formalism introduces coordinates for all possible wrapped branes. As such the duality groups are more complicated and the generalized metric becomes much more complicated. This will not effect the boundary issues we are discussing and so we will illustrate the bondary action for the case where the duality group is $SL(5)$ and we have only four dimensions involved in the duality transformation. In four dimensions only the membranes can wrap. The dual coordinates may be labelled by $y_{ab}$ (in four dimensions there are six such directions leading to a total generalized space of ten dimensions). The coordinates will be in the {\bf{10}} of SL(5) and the generalized metric will tranform linearly under SL(5) transformations. The coordinates of generalized space are denoted by:
\begin{equation}
 X^M=
\begin{bmatrix}
  x^a \\
  y_{ab}
\end{bmatrix}.
\end{equation}
(Note, the diffference in notation as compared with \eqref{2.3}. This is so as to be consistent in each case with the original literature).
A metric on this space can be constructed which will combine the ordinary metric and $C$ field, \cite{Berman:2010is} as follows. 
\begin{equation}
\begin{split}
 &M_{MN}=\left[\begin{array}{ccc}
	      g_{ab}+\dfr12 C_a^{\ph{a}ef}C_{bef} &	& \dfr{1}{\sqrt{2}}C_{a}^{\ph{a}kl} \\
	      &&\\
	      \dfr{1}{\sqrt{2}}C_{b}^{\ph{b}mn}	&	& g^{mn,kl}
            \end{array}\right];\\
\end{split}
\end{equation}
where $g^{kl,rs}=\fr12(g^{kr}g^{ls}-g^{ks}g^{lr})$.
The action in terms of this metric is gven by:
\begin{equation}
 \begin{split}
\label{1}
    V=&\sqrt{g}\left[\fr{1}{12}M^{MN}(\dt_M{}M_{KL})(\dt_N{M^{KL}})-\fr12{}M^{MN}(\dt_N{}M^{KL})(\dt_L M_{MK})\right.+\\
      &\left.+\fr{1}{12}M^{MN}(M^{KL}\dt_M M_{KL})(M^{RS}\dt_N M_{RS})+\fr14 M^{MN}M^{PQ}(M^{RS}\dt_P M_{RS})(\dt_M M_{NQ})\right],
 \end{split}
\end{equation}
where $\dt_M=(\fr{\dt}{\dt x^a},\fr{\dt}{\dt y^{ab}})$ and $M$ is the generalized metric.

As before, if one evaluates this action under the assumption that the fields are independent of the dual coordinates ie. $\partial_{y_{ab}}  =0$ then one recovers the usual Einstein Hilbert action with the kinetic term for the $C$ field.

The necessary boundary term is similar to the string theory cases. It again has the form of \eqref{2.4} but without the dilaton and the normal is again restricted to obey:
\begin{equation}
\label{9}
N_M=\left[\begin{array}{c}
            n_m\\ \\
	    0
           \end{array}
\right],\quad
 N^M=\left[\begin{array}{c}
            n^m\\ \\
	    -\dfr{1}{\sqrt{2}}C^n_{\ph{n}rs}n_n
           \end{array}
\right].
\end{equation}

This restriction of the normal is where one has chosen the fields to be independent of the $y$ coordinates. The general duality invariant restriction on the normal is not known since this is tied up with the physical section condition. Despite some obvious conjectures the physical section condition for the M-theory extended geometry is not known.

\section{Discussion}
 
The motivation for including these boundary terms is not just to provide a more thorough treatment of the duality reformulated actions but importantly to include the Gibbons-Hawking boundary term in the reformulated actions. The point then is to allow the thermodynamic properties to be calculated from these duality manifest actions through evaluating the action on shell. The hope would then be to use this to derive duality manifest thermodynamic properties of supergravity solutions. 

The other application of these surface terms is in constructing globally well defined solutions. These boundary terms provide important constraints on the construction of any topologically nontrivial solutions; the importance of such solutons in doubled geometry has been stressed recently in \cite{Andriot:2011uh}.

\section{Acknowledgements}
We wish to acknowledge discussions with Hiro Funikoshi, Gary Gibbons, Hadi and Mahdi Godazgar and Peter West.
DSB is partially funded by STFC rolling grant ST/J000469/1 and 
MJP by STFC rolling grant ST/J000434/1.

\bibliography{boundary}

\end{document}